# A Service-Based Architecture for enabling UAV enhanced Network Services

Oussama Bekkouche*, Konstantinos Samdanis‡, Miloud Bagaa*, Tarik Taleb*,⊗,1
*Aalto University, Espoo, Finland
‡ Nokia Bell Labs, Munich, Germany
⊗ University of Oulu, Oulu, Finland
1 Sejong University, Seoul, South Korea
emails: oussama.bekkouche@aalto.fi, konstantinos.samdanis@nokia-bell-labs.com,
miloud.bagaa@aalto.fi, tarik.taleb@aalto.fi

*Abstract*—This paper provides an overview of enhanced network services, while emphasizing on the role of Unmanned Aerial Vehicles (UAVs) as core network equipment with radio and backhaul capabilities. Initially, we elaborate the various deployment options, focusing on UAVs as airborne radio, backhaul and core network equipment, pointing out the benefits and limitations. We then analyze the required enhancements in the Service-Based Architecture (SBA) to support UAV services including UAV navigation and air traffic management, weather forecasting and UAV connectivity management. The use of airborne UAVs network services is assessed via qualitative means, considering the impact on vehicular applications. Finally, an evaluation has been conducted via a testbed implementation, to explore the performance of UAVs as edge cloud nodes, hosting an Aerial Control System (ACS) function responsible for the control and orchestration of a UAV fleet.

## I. INTRODUCTION

The fifth generation of mobile communications (5G) is expected to enable a plethora of new services including critical and disaster recovery communications. These services rely on stringent performance requirements in terms of latency, data rates and reliability. 5G aims to compel the user experience, by offering a consistent network performance throughout urban and rural environments. This is currently challenging, since factors like radio coverage variations due to obstacles and terrain structure, high user density and mobility may significantly impact network service capabilities.

To avoid costly infrastructure investments, which typically accommodate a fixed amount of traffic, UAV can offer flexible radio and networking extensions [1]. Indeed, UAVs can deliver temporary and disaster recovery networking and can follow a group of users, e.g. high speed vehicles, assuring service continuity[1]. The main advantage of UAVs is the ability to establish Line-of-Sight (LoS) communications, which assures a consistent connectivity with high channel quality [2]. Hence, there is a growing industry interest for integrating Unmanned Aerial Systems (UASs) into the mobile network ecosystem.

3GPP has introduced a reference model for UAS [3] as illustrated in Fig.1, wherein a UAV (treated as end-user device)

---

[1]GSMA Internet of Things, Accelerating the Commercial Drones Market using Cellular Connectivity, Nov. 2017

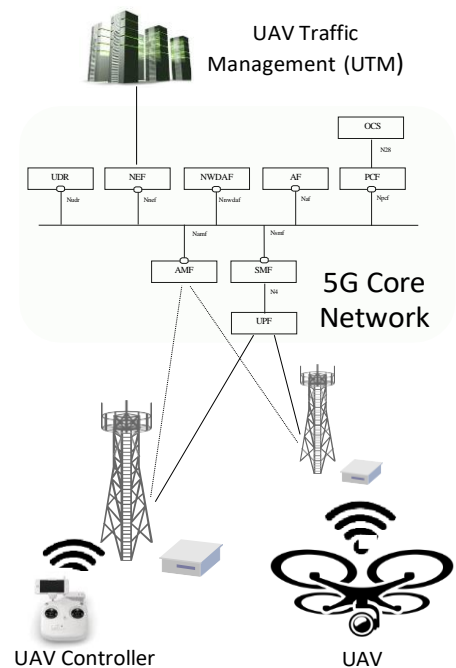

Fig. 1: The UAS reference model.

can be remotely controlled by a UAV controller either directly or via the network, i.e. enabling Beyond Visual Line-of-Sight (BVLOS) control. UAVs can then leverage the benefits of the 3GPP system in terms of coverage, authorization, tracking and Quality of Service (QoS) support. In addition, the 3GPP system enables the UAS Traffic Management (UTM) to interact with the UAS and allows authorized tenants, e.g. public safety agencies, to query the UAV identity and UAS meta-data [4].

Network services can use UAVs as aerial network equipment enhancing the performance of 5G systems. UAVs can provide a flexible coverage and capacity by adjusting their position; hence offering a means of "physical" network programmability. UAVs can also hold control plane, orchestration and edge cloud capabilities on demand, enabling scalable proximity services, Ultra-Reliable Low-Latency Communications (URLLC) and stable connectivity for a group of users. However, the computation resources on-board UAVs are limited due to their size, weight and power, which impose challenges on adopting conventional Network Functions (NFs) or added value services, driving the exploration of light-weight versions.

A preliminary approach of an airborne core network referred to as SkyCore based on Evolved Packet Core (EPC) is introduced in [5]. SkyCore allows the EPC to be collapsed into a single light-weight self-contented entity on-board each UAV, with inter-UAV connectivity control via Software Defined Networking (SDN).

This paper investigates the role of UAVs as 5G core network equipment with radio and backhaul capabilities and sheds light on the network architecture, introducing control, management and user plane enhancements. The contributions of this paper include: *i*) a study on UAV-based radio and core NFs; *ii*) shaping the Service-Based Architecture (SBA) for accommodating UAV aerial navigation, airborne user/control plane NFs, connectivity management and control; *iii*) elaborating on the impact of UAV enabled user/control plane capabilities for vehicular applications; and *iv*) analyzing via testbed implementations the performance of UAVs as edge cloud nodes hosting an Aerial Control System (ACS) function responsible for the control and orchestration of a UAV fleet.

The remaining of this paper is organized as follows. The next section overviews the use of UAVs in enhancing 5G network services. Section III elaborates on the role of UAVs as radio, backhaul, core network and edge cloud equipment, while Section IV describes the SBA enhancements for facilitating UAV-based NFs, considering also the network orchestration aspects. In Section V, we analyze the use of airborne user/control plane NF for enabling vehicular applications, and provide a testbed and simulation evaluation demonstrating: *i*) the UAV performance as edge cloud hosting an ACS network function, and *ii*) the UAV suitability for supporting URLLC. Finally, Section VI concludes the paper.

## II. UAV ENHANCED NETWORK SERVICES

UAVs can offer network services closer to end-users with a higher quality and that is thanks to their stronger reference signals, due in turn to clear LoS. The coverage of UAVs depends on the altitude, antenna characteristics including directivity and transmission power, as well as on weather conditions. Higher altitudes enable the detection of base stations at further distances, e.g. at 400ft a UAV can connect up to 18 base stations depending on the frequency band[2]. Typically UAVs operate as a fleet, which distribute service responsibilities. A UAV fleet can provide complex tasks, combining communication and optionally value added services, and can last longer balancing energy expenditures. With respect to 5G, UAVs can facilitate the following enhancements:

[2]Qualcomm, Leading the World to 5G: Evolving Cellular Technologies for Safer Drone Operation, Sep. 2016.

- **Broadband coverage** on a temporary basis, e.g. for an event, or compensate coverage holes in case of disaster on a short time span.
- **Enhanced capacity,** which relaxes network over-provisioning, by enabling temporary services, e.g. at peak times, and offer data offloading via a local cloud.
- **Deterministic behavior,** which enhances the data plane resiliency, by enabling diverse disjoint paths, while ensuring low latency through better propagation conditions.
- **Firm communications for high speed vehicles,** which provides a stable control/user plane connectivity taking advantage of the UAV's altitude and LoS.
- **Intelligent services for critical applications**, e.g. sensors and cameras for security, local cloud processing for object or face recognition, etc.

UAVs are obliged to follow certain predetermined airway routes, with planned intersections and parking positions [6]. Such routes information can ease control and network integration for UAVs [7], while assuring public safety. The ownership of UAVs impacts service orchestration. When UAVs are part of the mobile operator infrastructure, the control is simpler. Whilst when UAVs belong to a vertical, their use requires authentication and synchronization with the network.

## III. UAV AS A NETWORK EQUIPMENT

The role of UAVs on enhancing 5G networks can be widely diverse, with UAVs acting as radio, core network, edge cloud or fronthaul/backhaul equipment. Since UAVs have energy limitations, considerations should be made for the travel plan as well as for the usage of computational, storage and communications capabilities.

### A. UAVs with Radio and Transport Layer Capabilities

5G radio, aka New Radio (NR), offers advanced features, including: *i*) service-tailored operations, *ii*) functional split into a Central Unit (CU) and one or more Distributed Units (DUs) allowing control and data plane separation, and *iii*) a new lower layer enabling a higher spectrum. 5G Radio Access Networks (RANs) are enriched to support an Integrated Access and Backhaul (IAB) [8], wherein CU and DUs are connected via a multi-hop wireless medium. Initially, the 5G RAN deployment is expected to be spotty and hence the co-existence with other radio technologies is essential. UAVs can support such RAN co-existence [9] offering on-demand the following radio capabilities:

- **Flying RAN station**:
  1) *In band radio relay*, that forwards (or processes and forwards) data traffic towards a donor macro cell.
  2) *Small cell or macro base station* with a wireless backhaul towards a gateway or macro base station on the ground.
  3) *Remote Radio Head (RRH)* connected to a Base Band Unit (BBU) on the ground. Different functional splits can be supported depending on connectivity quality and cloud capabilities on UAVs.
  4) *WiFi access* using unlicensed spectrum.

5) *Delay Tolerant Network (DTN) node*, which can store and forward data opportunistically assisting delay tolerant applications.
- **Flying 5G RAN component**:
    1) *CU or DU* with lower layer capabilities used for access node redundancy and path diversity with wireless connectivity.
    2) *NR Node B (gNB) or millimeter Wave (mmWave)* offering high speed data rates.
    3) *Control plane* of gNB or mmWave access.
    4) *x-haul node*, a non-ideal backhaul node with heterogeneous transmission links, which forms the IAB architecture.
- **Flying local services**:
    1) *Device-to-device (D2D) node* providing proximity or location services.
    2) *Aerial Road Side Unit (ARSU)* for assisting advanced vehicular communications.

UAV enabled wireless links and interference control differ compared to conventional radio communications. The reason is the strong dependency on weather conditions (e.g., fog and rain) and on LoS, which relies on UAV location and on the antenna elevation of ground macro cells. UAVs in close proximity can also suffer from co-channel interference and hence cooperation techniques are essential. Mobility may raise the complexity of interference control and can potentially create Doppler shift that causes severe inter-carrier interference, especially for mmWave. A UAV with radio capabilities is also required, with the exception of in-band relays, to support a wireless transport connectivity towards a macro or gateway node on the ground. Alternatively, UAVs can also act as backhaul nodes facilitating, e.g. micro-wave or mmWave links. UAV mobility impacts the network topology and routing protocols as documented in [10], affecting single-hop and multi-hop connectivity.

*B. UAVs carrying Core Network Functions*

UAVs may hold core NFs, assuming that the radio and transport layer capabilities are also present. NFs residing at UAVs can complement the core network on the ground, facilitating cooperation. It is practical to hold core NFs at a UAV when considering a fleet, offering resiliency, alternative routing and allowing flexible connectivity towards the network on the ground. Core NFs on UAVs can reduce latency while ensuring stable connectivity for mobile users. UAVs can also provide a means of collecting UE and local environment data for enriching network analytics and automation, e.g. for root cause analysis and QoS assurance. Since computation resources on-board UAVs are limited, it is challenging to run virtual machines, selecting instead the adoption of containers. Our testbed evaluation study provides insightful results about the ability of UAVs to host container-based NFs. Given the fact that UAVs have energy limitations, core NFs that hold user registration data, policy and authentication are not suitable to be hosted on UAVs to avoid losing critical data that may disrupt other operations. Typical core NFs that can be hosted at UAVs may include:

- **Control plane functions**:
    1) *Access and Mobility Function (AMF)* can regulate connectivity and reachability closer to the user, while providing an effective location service management.
    2) *Session Management Function (SMF)* selects User Plane Functions (UPFs) hosted on UAVs and controls efficiently traffic steering. An airborne SMF can determine QoS rules and the Session and Service Continuity (SSC) mode towards edge or Local Data Networks (DNs) carried on UAVs or on the ground.
    3) *Network Exposure Function (NEF)* enables third parties to control UAV operations and may introduce content and services on UAV hosted Local DNs.
    4) *Network Data Analytics Function (NWDAF)* in conjunction with management data analytics can exploit sensors and terrain monitoring capabilities on UAVs collecting advanced data for network load, interference control and coverage, mobility and maintaining LoS.
- **User plane and edge cloud**:
    1) *Local UPF* introduces a distributed user plane node on selected UAVs, which enforces policy rules, while reporting traffic usage. A local airborne UPF can enable route optimization and multi-hop data forwarding exploiting route diversity toward the mobile network on the ground. It also facilitates access on Local DN applications, hosted on the same or another UAV.
    2) *Edge cloud* computing allows content, applications and proximity services or simply computing to be offered by UAVs provided that a local UPF facilitates connectivity and SSC services.

Depending on the deployment scenario, selected control and user plane functions or even the edge cloud can be co-located. The benefits of co-locating NFs and cloud capabilities on a single UAV include improvements on service latency and signaling overhead, while the limitations concentrate mainly on energy expenditure and reliability. To enable the selection of NFs on UAVs, there is a need to enhance the Network Repository Function (NRF) to indicate that a particular NF is held on a UAV considering for instance location and height. Since NFs on UAVs should use limited resources, while minimizing energy expenditures, the adoption of a proxy-based or light-weight NF with selected functionality may prove to be beneficial. To simplify the operation of NFs on UAVs, a pre-configured policy can reduce the interaction with the core network. Such a policy can provide rules for determining the point of attachment, SSC mode, traffic steering, etc., based on expected performance, network status and UAV travel plan.

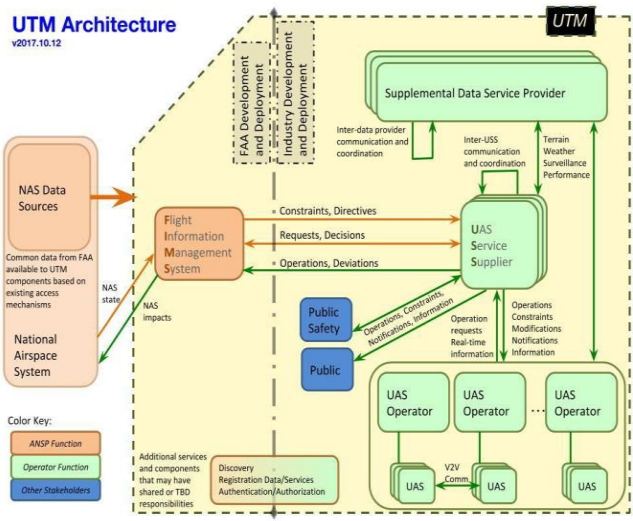

Fig. 2: UTM architecture [11].

IV. SBA ENHANCEMENTS FOR SUPPORTING UAV NETWORK SERVICES

To efficiently support UAV-based network services, there is a need for enabling a consistent cooperation with aerial navigation considering also weather conditions. This can be achieved by allowing an interaction among the core network with navigation and air traffic management, i.e. UTM, introducing control and orchestration enhancements for handling UAV cooperation and connectivity.

A. UAS Traffic Management (UTM) System

UTM systems enable Aviation Safety Agencies (ASAs), such as the Federal Aviation Administration (FAA) and European Aviation Safety Agency (EASA), to support UAS aerial navigation [11]. UTM's components may belong to distinct stakeholders including the UAS operator, Air Navigation Service Provider (ANSP) and other third parties as illustrated in Fig. 2. The use of UTM considers distributed services, which may interact via Application Programming Interfaces (APIs) allowing third parties to provide customized control. The UTM components and the respective stakeholders may include:

- **ANSP** that provides a regulatory and oversight framework to airspace operators, by exchanging information related to airspace restrictions, while accessing flight data at any time. The most important service of ANSP is the *Flight Information Management System (FIMS)*, which acts as a communication gateway towards other UTM services.
- **Ancillary stakeholders** are third parties such as public safety and government agencies that access and/or provide UTM services to ensure security and privacy or law enforcement related to UAVs' flights.
- **Supplemental Data Service Provider (SDSP)**, which offers additional information to the UAS operator, such as weather forecasts, terrain and obstacle data.
- **UAVs operator** can optionally own the UAVs and/or the UAS controller and is responsible for flight safety by respecting regulatory instructions, resolving conflicting flight plans and sharing flight information. The main related services are:
    1) *UAS Service Supplier (USS)* acts as a communication gateway between the operator and other UTM's actors. The USS assists the operator for flight planning, conflict resolution and conformance monitoring. It also maintains statistics about airspace traffic for analytics, regulatory and accountability purposes.
    2) *UAS Controller* collects flight information from USS and SDSP to control the UAVs flights. Depending on the operation mode of UAVs i.e. standalone or fleet, it may be deployed as a *Ground Control System (GCS)* and/or an *Aerial Control System (ACS)* that runs on-board an elected UAV [12]. In case both controllers are present, the GCS may control a number of ACSs.

For UAV-based network services, the mobile network operator may play the role of the UAV operator and may introduce new procedures and functions that are crucial for supporting specific applications.

B. SBA enhancements for UAV Navigation and Control

The support of UAV-based network services can be simplified once the UTM architecture is integrated with the core network. This is feasible given the fact that a UTM system consists of a set of distributed and federated services that communicate via APIs [11], which are aligned with the main principles of SBA. Indeed, each UTM service integrated into the SBA is self-contained and cloud-native.

Certain enhancements are needed for handling connectivity, navigation and control. Such UAV services aim to complement coverage, networking and added value services offered by the mobile network operator or third party. The UAS controller, i.e. GCS and/or ACS, and USS are essential elements for controlling and planning the UAVs flight, and hence need to interact with other NFs to enable efficient resource management and stable connectivity with the desired QoS via the SBA.

SDSP, FMIS and ancillary stakeholders functions hold UAV related application details, i.e. can act as Application Functions (AFs), which interact with the operator network via the means of USS. Provided that the USS resides inside the operator network, authentication and information exchange with third parties can be facilitated via the NEF. Besides the input collected by USS, UAS controllers, i.e. GCS and ACS, require also assistance regarding network connectivity and resource management. To accomplish this, a new control plane NF named *UAV-based Network Service Control (UNSC)* is introduced to handle connectivity, resource management

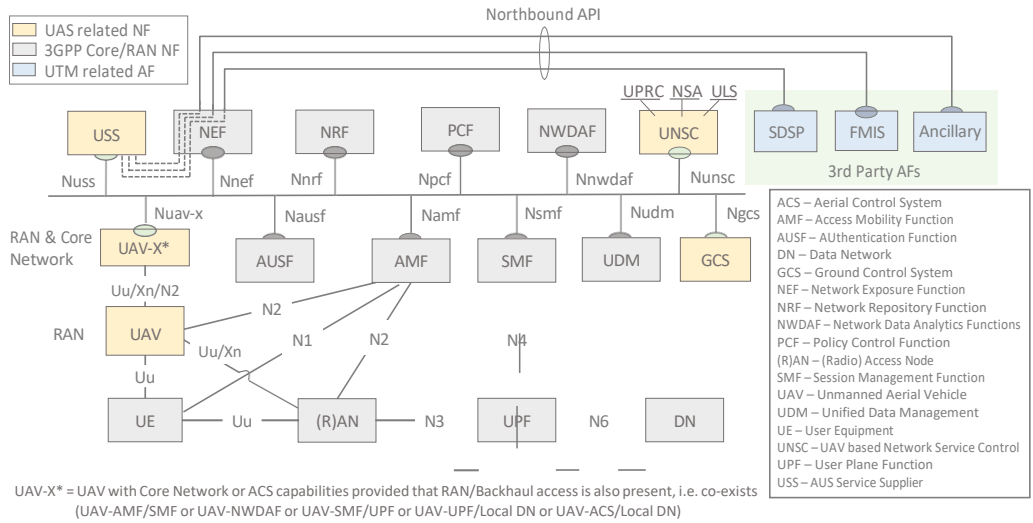

Fig. 3: Enhanced SBA supporting UAV Network Services.

and location services related to UAVs. The operation of UNSC relies on the following sub-functions:

- **UAV Policy and Resource Control (UPRC)** provides assistance to the corresponding GCS and/or ACS regarding the UAV connectivity in terms of SSC and preferences towards the network on the ground.
- **Network Service Assurance (NSA)** assists the GCS and/or ACS to configure and maintain a constant UAV network service despite energy limitations. NSA keeps track of the UAV energy and provides suggestions for replacing energy limited UAVs or altering the topology and connectivity, based on the flight plan and parking locations similarly to [13].
- **UAV Location Services (ULS)** assists the network to estimate the location of UAVs without GPS on-board that conserves battery, based on 5G Location Services (LCS). ULS reports the location of UAVs, similarly as UEs, to GCS and/or ACS.

An overview of an enhanced SBA, capable to accommodate UAV-based network services, is illustrated in Fig.3, where the newly introduced NFs and AFs related to UAS control and UTM components are shown alongside the mobile core network. UAVs can hold control plane NFs, e.g. UAV-AMF/SMF or UAV-NWDAF, or a combination of control and user plane NFs, such as UAV-SMF/UPF. Alternatively, a UAV may act as an edge cloud supporting value added services, via the means of UAV-UPF/Local DN with optionally including NEF for allowing third parties to interact and configure cloud services. A UAV hosting an edge cloud can also hold an ACS functionality, i.e. UAV-ACS/Local DN, for controlling UAVs directly or in coordination with GCS.

Since UAV network services are sensitive to radio, weather and positioning conditions that influence connectivity and QoS performance, it is beneficial to exploit analytics, i.e. NWDAF, to feed the UNSC with information related to the expected network load at specified areas of interest. Such analytics can assist UNSC to improve traffic steering decisions, UAV positioning adjustments and connectivity towards the mobile network on the ground. The adoption of NWDAF in conjunction with management data analytics on UAVs can further provide terrain information and UAV-related analytics, e.g. QoS performance and energy consumption, to the UNSC.

UNSC should coordinate with the GCS and/or UAV-ACS for selecting NFs on-board UAVs, maintaining connectivity, executing resource adjustment and traffic steering. In addition, UNSC interacts with the management and orchestration plane for triggering resource scaling and/or NF re-location. This is feasible since SBA, as an architecture model, is also adopted in the management plane allowing an easier interaction with the core network [14].

*C. Orchestration and Control of UAV enhanced 5G Networks*

UAVs operating on 5G networks should accommodate network services in both spatial and temporal dimensions. Hence, active monitoring and corrective measures are needed, besides planning, to assure that the appropriate connectivity is established. To address such requirements, the mobile network needs to facilitate the following two fundamental functionalities:

- **Network orchestration** supports resource planning taking into account the entire UAV travel plan, including speed and height. The goal is to allocate the desired amount and type of resources, i.e. radio, cloud and backhaul, which fulfill particular service characteristics, e.g. capacity, latency, etc., determining also the appropriate attachment points on the ground for the duration of the flight [15].
- **Resource adjustment** is triggered upon a UNCS request, which provides to the orchestrator the adjusted flight route, taking into account weather and radio conditions. Corrective measures are then provided to the UNCS reflecting UAVs re-positioning, which can include scale-up/down or relocation of network and cloud resources, considering new attachment points towards the mobile network on the ground.

As the allocation of network resources is based on a predetermined travel plan, it is possible to estimate the time and duration during which radio, cloud and backhaul resources become needed at every point of attachment, respectively. The orchestrator should take into account the expected UAV residual time considering speed and altitude to enable an efficient resource allocation. The resource planning process should also reflect potential deviations in UAV's location and the respective connectivity adjustments for introducing network service robustness, which can assist: *i*) energy limited UAVs, *ii*) alternating traffic conditions and *iii*) dynamic third party requests. Resource adjustment procedures may rely on analytics that can predict and assess potential resource adjustments, while interacting with the orchestrator.

## V. EVALUATING THE IMPACT OF UAV-ASSISTED NETWORK SERVICES

This section contains an evaluation study related with UAV-assisted network services. The first part analyzes the use of UAVs to enable network services across the RAN, core network and backhaul for supporting various vehicular-to-everything (V2X) applications considering a high mobility environment that brings challenges in service continuity, latency and reliability. Meanwhile, the experimentation analysis, presented in the second part, provides insightful results related to the use of UAVs as edge cloud, evaluating the potential of hosting an ACS function or any other core NF.

### A. A Qualitative Analysis of UAV-assisted V2X Services

To elaborate the impact of UAV-based network services on high speed users, we consider the scenario of V2X, wherein UAVs follow selected vehicles in close proximity. Fig.4 illustrates the UAV-assisted network services, which impact the RAN or the entire network, including the backhaul (BH) and core network, with respect to V2X related applications. UAV-assisted RAN services can extend the coverage and/or capacity of Ground RAN (G-RAN) using licensed (relay) or unlicensed (WiFi) spectrum, delivering best effort services or DTN enabled portable data offloading applications. Similarly, aerial Small Cells (SC) or Base Stations (BS) can extend the G-RAN coverage and capacity, offering 5G QoS services. The optional user and control plane (UP/CP) split, i.e. introducing flying phantom SC/BS [16], can enhance spectrum control and offer robustness against high mobility.

For UAV-assisted network services, combining aerial SC/BC with wireless BH capabilities and core NFs, can enable the following vehicular applications:

- AMF/SMF can connect vehicles not in vicinity, offering a bird eye view, while enabling a faster and more stable integration with the core network.
- A combination of AMF/SMF and UPF/Local DN can assure low-latency edge cloud services due to direct LoS communications enhancing the offered capacity. A moving cloud service can also cope effectively with vehicle mobility, while the optional support of NEF allows third party interaction with aerial cloud applications.
- Coupling AMF/SMF with UPF/NWDAF, considering an arrangement that employs multiple UPFs, can offer low-latency and enhanced reliability by introducing alternative paths with LoS via airborne UPFs. The use of NWDAF can provide analytics for assessing the load and connectivity quality considering the terrain constraints and vehicles' mobility.

The optional presence of additional wireless airborne BH nodes, in between the BS/SC and core network ones, can assure LoS, e.g. in urban canyons, provided that the delay introduced by the wireless BH medium is limited. An aerial SC/BS can also offer proximity services or enable a D2D sidelink, being responsible for efficient spectrum allocation among vehicles, while it may also provide ARSU services for enhancing driving experience and safety. Aerial SC/BS(s) can extend further the G-RAN coverage at specified regions, once utilizing additional wireless BH capabilities offered by other neighboring UAVs. Such an arrangement mostly facilitates best-effort services, since delay increases and capacity deteriorates in a wireless multi-hop environment.

UAV-assisted cloud-RAN services employ a set of RRHs aboard a fleet of UAVs, which are connected via an x-haul (XH) wireless medium to a CU/DU or directly to the G-RAN. Such an arrangement enables a virtual BS that offers a stable connectivity for highly mobile vehicles. Supporting AMF/SMF and UPF/Local DN(NEF) capabilities on UAVs that hold a CU/DU functionality can additionally enable cloud services on the move, enhancing the G-RAN capacity and service stability.

### B. A Testbed Study on UAV-based ACS NFs

This section demonstrates the performance of UAVs that host lightweight virtualized NFs in an enhanced SBA environment. In our experimentation, we consider a fleet of UAVs, where a UAV is assigned as an airborne edge cloud platform that hosts a set of UAV-ACS instances responsible for monitoring and controlling the UAV fleet by exchanging telemetry data, e.g. speed, GPS position, current heading and battery level, as well as Command and Control (C&C). The developed airborne edge cloud platform, in our testbed, is compliant with ETSI Multi-access Edge Computing (MEC) architecture [17] and consists of the following components:

- **MEC platform** based on LXD, a next generation system container manager. The developed MEC platform allows life-cycle management of MEC applications (i.e., creation, deletion and reallocation), management of traffic rules and capability exposure via RESTfull APIs.
- **UAV-ACS** that is implemented as a MEC application. Each UAV-ACS instance runs inside a dedicated LXC container and is responsible for the control and monitoring of UAVs that belong to the fleet, using the telemetry protocol **MAVLink**. A UAV-ACS instance exposes a REST API that allows interaction with other NFs, e.g. with UNSC for updating the flight path, getting the current location and communicating the battery level.
- **Virtualisation infrastructure** including the compute, storage, and network resources provided by a companion computer on-board UAVs. Our testbed adopted the *Intel UP Squared* as a companion computer, which achieves a good balance between performance and size. It is equipped with Intel Atom X7-E3950 processor (quad core), 8GB RAM, and 64GB eMMC, with a dimension of 3.37" x 3.54" and with a 802.11n WI-FI dongle used for communication with other UAVs.

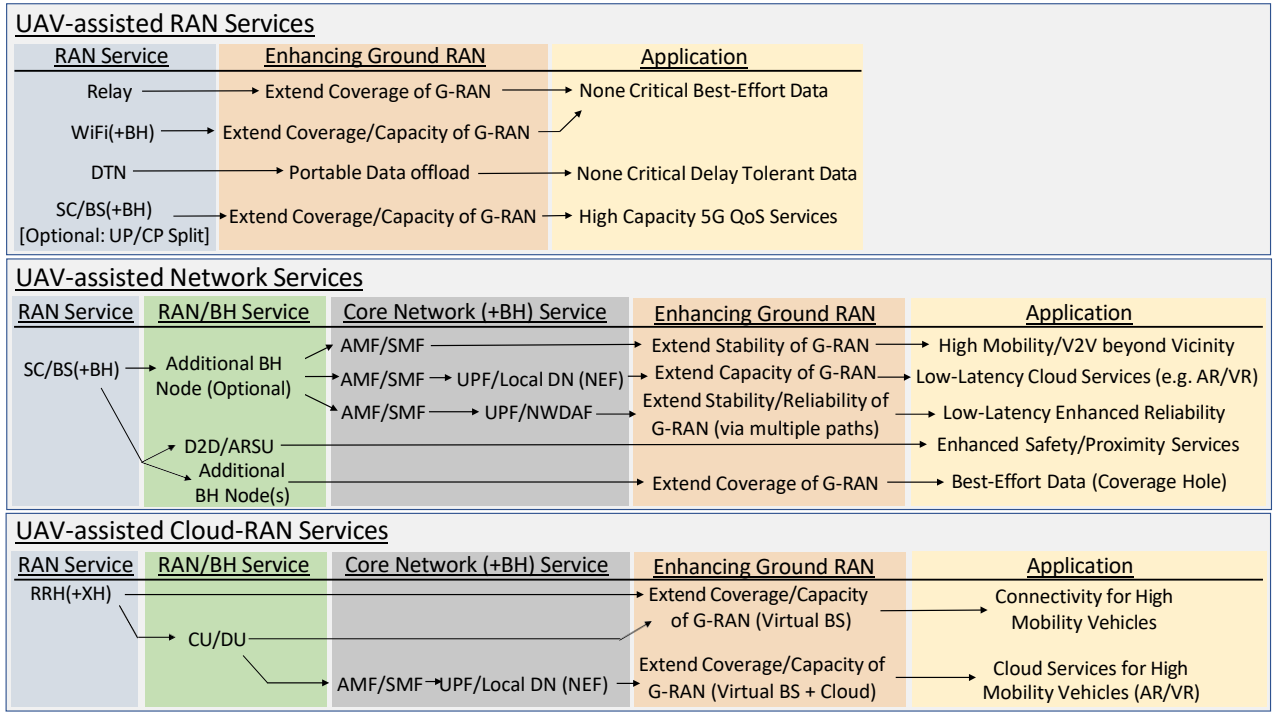

Fig. 4: UAV-assisted Network Services for enhancing Vehicular-to-Everything.

Table I summarizes a set of average performance measurements related to the deployment of UAV-ACS instances on-board UAVs. The deployment time shows the time required to instantiate and launch a new instance of the virtualized UAV-ACS. The reallocation time reflects the complexity related with the process that impacts the time required for reallocating a running instance of UAV-ACS to a new airborne edge cloud host, i.e. from a UAV to another, by means of live migration using the LXD embedded live migration tool. Service reallocation is of crucial importance for UAVs-based services due to high mobility and energy constraints. A related measure is the downtime that represents the portion of time during which the service is not available and that is only around 12% of the overall reallocation time. The communication performance among two UAVs is also considered in terms of the maximum available bandwidth, communication latency, and packet loss with the obtained results confirming that UAVs are suitable for hosting services with stringent latency requirements.

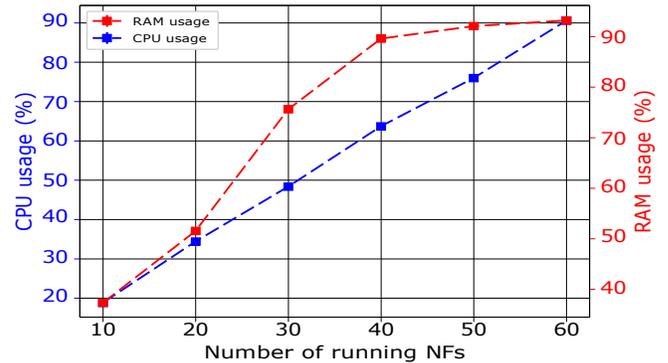

Fig. 5: UAV-ACS NFs resources consumption.

TABLE I: UAV-based network service performance.

| Metric | Average value |
|---|---|
| Deployment time | 3.894(s) |
| Reallocation time | 51.963(s) |
| Downtime | 6.415(s) |
| Available bandwidth | 12.44Mbits/s |
| Latency (RTT) | 1.276(ms) |
| Packet loss | 0.0437% |

To demonstrate the scalability of the UAV-based edge cloud platform, the UAV's resource consumption is measured in terms of CPU and RAM, considering different numbers of running UAV-ACS instances for the control of the fleet. To emulate a large fleet of UAVs, a customized version of the open-source SITL (Software In The Loop) emulator is used that allows running the open-source UAVs' autopilot ArduCopter in a normal PC without any dedicated hardware and with the same behavior as a real UAV. Such customized version establishes a telemetry connection towards a UAV-ACS instance and supports its reallocation to another UAV.

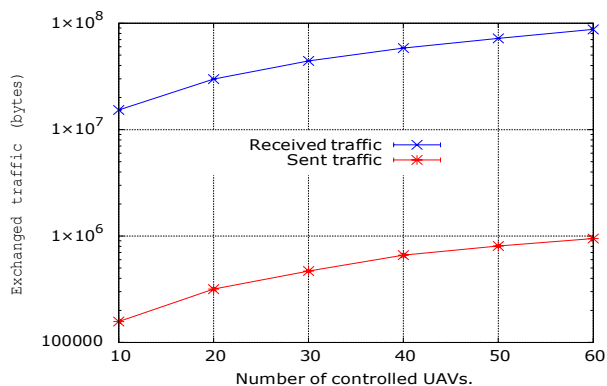

Fig. 6: UAV-ACS control message exchange scalability.

Fig. 5 depicts the usage evolution of CPU and RAM as a function of the number of UAV-ACS instances. The CPU usage is observed to linearly increase along with the number of UAV-ACS instances, until it reaches a total usage of $90\%$ for $60$ instances. The extracted slope shows that each instance consumes around $1.43\%$ of total CPU time. The RAM usage follows a logarithmic evolution where it reaches $93\%$ of the total available RAM for $60$ instances. In our experimentation, it is noticed that the maximum number of UAV-ACS instances on-board a UAV cannot surpass $60$, otherwise some telemetry connections are dropped.

To provide an insight regarding the traffic amount exchanged during the control and monitoring process of UAVs, the telemetry traffic generated by the UAV-ACS instances is measured during a period of $3$ minutes. The results plotted in Fig.6 using a logarithmic scale allow us to observe the evolution of sent/received traffic during monitoring and control in the same graph. Both curves are nearly linear with the amount of traffic sent being negligible compared to the one received. Indeed, the total amount of received traffic increases from $15.3$MB when the number of controlled UAVs is $10$ to $87.5$MB for $60$ UAVs, while the amount of sent traffic increases from $156.8$KB to $944$KB for $10$ to $60$ UAVs, respectively.

## VI. CONCLUSION

This paper explores the role of UAVs in facilitating core NFs with radio and backhaul capabilities. It analyzes the required advancements in SBA, considering air traffic management, weather conditions and network connectivity and introduces new resource control and management functions to efficiently deal with topology and service assurance. The use of UAVs for enhancing 5G network services is then evaluated and shown to be useful considering a range of V2X applications. A testbed and simulation analysis shed light onto the performance of UAVs as edge clouds hosting ACS instances responsible for the control and orchestration of a UAV fleet.

ACKNOWLEDGMENT

This work was supported in part by the European Unions Horizon 2020 programme under the EU 5G!Drones project (Grant No. 857031) and in part by the Academy of Finland 6Genesis Flagship.

REFERENCES

[1] B. Li, Z. Fei, and Y. Zhang, "UAV communications for 5G and beyond: Recent advances and future trends," *IEEE Internet of Things Journal*, vol. 6, no. 2, pp. 2241–2263, Apr. 2019.
[2] H. Hellaoui, A. Chelli, M. Bagaa, and T. Taleb, "Efficient steering mechanism for mobile network-enabled uavs," in *2019 IEEE Global Communications Conference (GLOBECOM)*, 2019, pp. 1–6.
[3] 3GPP TR 22.825, "Study on remote identification of unmanned aerial systems (UAS)," Rel.16, Sep. 2018.
[4] A. Fotouhi, H. Qiang, M. Ding, M. Hassan, L. Giordano, A. Garcia-Rodriguez, and J. Yuan, "Survey on UAV cellular communications: Practical aspects, standardization advancements, regulation, and security challenges," *IEEE Communications Surveys and Tutorials*, vol. 21, no. 4, pp. 3417–3442, 4th Quarter 2019.
[5] M. Moradi, K. Sundaresan, E. Chai, S. Rangarajan, and Z. Mao, "SkyCore: Moving core to the edge for untethered and reliable UAV-based LTE networks," *ACM MobiCom*, Oct.-Nov. 2018.
[6] M. Gharibi, R. Boutaba, and S. Waslander, "Internet of drones," *IEEE Access*, vol. 4, pp. 1148–1162, April 2016.
[7] O. Bekkouche, M. Bagaa, and T. Taleb, "Toward a utm-based service orchestration for uavs in mec-nfv environment," in *2019 IEEE Global Communications Conference (GLOBECOM)*, 2019, pp. 1–6.
[8] 3GPP TR 38.874, "Study on integrated access and backhaul," Rel.16, Jan. 2019.
[9] I. Bor-Yaliniz and H. Yanikomeroglu, "The new frontier in ran heterogeneity: Multi-tier drone-cells," *IEEE Communications Magazine*, vol. 54, no. 11, pp. 48–55, November 2016.
[10] J. Jiang and G. Han, "Routing protocols for unmanned aerial vehicles," *IEEE Communications Magazine*, vol. 56, no. 1, pp. 58–63, Jan. 2018.
[11] NASA UTM, "Unmanned Aircraft System (UAS) Traffic Management (UTM), UTM ConOps v1.0," May 2018.
[12] H. Hellaoui, O. Bekkouche, M. Bagaa, and T. Taleb, "Aerial control system for spectrum efficiency in UAV-to-cellular communications," *IEEE Communications Magazine*, vol. 56, no. 10, pp. 108–113, Oct. 2018.
[13] G. Secinti, P. Darian, B. Canberk, and K. Chowdhury, "SDN in the sky: Robust end-to-end connectivity for aerial vehicular networks," *IEEE Communications Magazine*, vol. 56, no. 1, pp. 16–21, Jan. 2018.
[14] 3GPP TS 28.532, "Management and orchestration; generic management services," Rel.15, Jan. 2020.
[15] O. Bekkouche, T. Taleb, M. Bagaa, and K. Samdanis, "Edge cloud resource-aware flight planning for unmanned aerial vehicles," *IEEE WCNC, Marrakech, Morocco*, Apr. 2019.
[16] DOCOMO, "5G radio access: Requirements, concept and technologies," *5G White Paper*, Jul. 2014.
[17] ETSI GS MEC 003, "Multi-access edge computing (MEC); framework and reference architecture," V2.1.1, Jan. 2019.

BIOGRAPHIES

**Oussama Bekkouche** is a doctoral student at the School of Electrical Engineering, Aalto University. He received his B.E. and M.Sc. from the University of Science and Technology Houari Boumediene (USTHB), Algeria, in 2015 and 2017, respectively. His research focuses on 5G networks, Unmanned Aerial Vehicles, and the Internet of Things.

**Konstantinos Samdanis** is a research project manager at Nokia Bell Labs Munich involved in standardization activities on 5G core networks. He was a visiting researcher at Aalto University and a principal researcher at Huawei German Research Center Munich, focusing on 5G carrier networks. Previously, he worked for NEC Europe Heidelberg as a senior researcher and a broad-band standardization specialist. He received his Ph.D. and M.Sc. degrees from Kings College London

**Miloud Bagaa** is currently a senior researcher at the School of Electrical Engineering, Aalto University, Finland. Prior to his current academic position, he was working as postdoctoral fellow with the European Research Consortium for Informatics and Mathematics, and he worked with the Norwegian University of Science and Technology, Trondheim, Norway. Before joining NTNU and till February. 2015, he worked as researcher at the Research Center on Scientific and Technical Information (CERIST), Algiers, Algeria.

**Tarik Taleb** is Professor at Aalto University and University of Oulu. He is the founder and director of the MOSA!C Lab (www.mosaic-lab.org). Prior to that, he was a senior researcher and 3GPP standards expert at NEC Europe Ltd., Germany. He also worked as assistant professor at Tohoku University, Japan. He received his B.E. degree in information engineering, and his M.Sc. and Ph.D. degrees in information sciences from Tohoku University in 2001, 2003, and 2005, respectively.